\documentclass[trackchanges,twocolumn]{aastex701}

\usepackage{float}

\begin{document}

\title{Multi-messengers from the radioactive decay of $r$-process nuclei}

\author[orcid=0000-0002-7893-4183]{Axel Gross}
\affiliation{Theoretical Division, Los Alamos National Laboratory, Los Alamos, NM 87545, USA}
\affiliation{ Center for Theoretical Astrophysics, Los Alamos National Laboratory, Los Alamos, NM 87545, USA}
\email[show]{agross@lanl.gov}  

\author[orcid=0000-0003-1758-8376]{Samuel Cupp}
\affiliation{Theoretical Division, Los Alamos National Laboratory, Los Alamos, NM 87545, USA}
\affiliation{ Center for Theoretical Astrophysics, Los Alamos National Laboratory, Los Alamos, NM 87545, USA}
\email{scupp@lanl.gov}  

\author[orcid=0000-0002-9950-9688]{Matthew R. Mumpower}
\affiliation{Theoretical Division, Los Alamos National Laboratory, Los Alamos, NM 87545, USA}
\affiliation{ Center for Theoretical Astrophysics, Los Alamos National Laboratory, Los Alamos, NM 87545, USA}
\email{mumpower@lanl.gov}

\begin{abstract}
The radioactive $\beta$-decay of nuclei synthesized in the rapid neutron capture process ($r$-process) releases a variety of particles, including electrons, $\gamma$-rays, neutrinos, and neutrons. 
These particles provide a rich set of multi-messenger signals that carry information about the astrophysical environments where neutron-rich nucleosynthesis occurs. 
In this work, we calculate from first principles the emission spectra resulting from the $\beta$-decay of $r$-process nuclei. 
Our approach incorporates detailed nuclear structure and decay data to model the energy distributions of each particle species. 
We couple the spectra with a nuclear reaction network simulation to obtain the temporal evolution of these distributions. 
We find that the emission distributions vary significantly in time and are non-thermal, with substantial average energies. 
We investigate these nuclear signals as a direct probe of heavy element formation and show that they are complementary observables to kilonova.  
\end{abstract}

\keywords{\uat{Gamma-ray bursts}{629}, 
\uat{Nuclear astrophysics}{1129}, 
\uat{Nucleosynthesis}{1131}, 
\uat{R-process}{1324}, 
\uat{Compact objects}{288}
}

\section{Introduction} 
It has long been proposed that the rapid neutron capture process (\textit{r}-process) is responsible for the synthesis of a significant fraction of the elements heavier than iron (see e.g., \cite{1957RvMP...29..547B} , \cite{2023A&ARv..31....1A}). 
The \textit{r}-process occurs in astrophysical environments with large numbers of free neutrons, and proposed sites include supernovae (see e.g. \citealt{1994ApJ...433..229W,1996ApJ...471..331Q,2001ApJ...554..578W,2006ApJ...646L.131F,2012ApJ...750L..22W,2015ApJ...810..109N,2019Natur.569..241S,2021Natur.595..223Y}) as well as compact object mergers, e.g. the collision of two neutron stars (NSMs) (see \citealt{1974ApJ...192L.145L,1989Natur.340..126E,1999A&A...341..499R,1999ApJ...525L.121F,2006ApJ...646L.131F,2011ApJ...738L..32G,2012MNRAS.426.1940K,PhysRevD.87.024001,2013ApJ...773...78B,2014ApJ...789L..39W}). Due to the extreme conditions present in these environments, as well as limited experimental data for many neutron-rich isotopes far from stability where the \textit{r}-process occurs, it is difficult to obtain precise predictions for \textit{r}-process production yields; and therefore, the relative contributions of these sites is not well understood. 

In recent years, there has been an increased focus on the use of astrophysical observations to provide constraints on the astrophysical location of the $r$-process. The landmark multi-messenger observations of GW170817 (see \cite{2017ApJ...848L..12A}) have given credence to NSMs as being a site of robust \textit{r}-process nucleosynthesis. This conclusion stems from the observed kilonova, a radioactive transient thought to be powered by the decay of $r$-process nuclei (see \citealt{2017Sci...358.1570D,2017Natur.551...80K,2017Sci...358.1559K,2017ApJ...848L..34M,2017PASJ...69..102T,2018MNRAS.481.3423W,Brethauer:2024zxg}). The production of lanthanide elements which have high opacity (e.g., \citealt{2013ApJ...774...25K,2013ApJ...775..113T}) has been shown to result in a spectrum that peaks in the infrared, and therefore, the observed optical and near-infrared peak was interpreted as evidence that NSMs undergo a strong $r$-process. In contrast, a weaker $r$-process is projected to yield a peak in the blue and optical bands (e.g., \citealt{2010MNRAS.406.2650M,2011ApJ...736L..21R,2014MNRAS.441.3444M}).

Although kilonova are of great interest and importance to the astrophysical community, the necessary multi-physics involved makes the modeling challenging. The numerous poorly constrained parameters in kilonova models can lead to degenerate light curves, obscuring the physical interpretation of these events (see e.g., \citealt{2016ApJ...829..110B,2021ApJ...906...94Z,2021ApJ...910..116K,Fryer:2023pew,Fryer:2023osz}). 

One principal component of modeling the kilonova light curve is the spectral emission profiles of radioactive decay products. The large number of nuclear $\beta$-decays which occur during the course of the $r$-process generates an enormous number of emitted particles: 
electrons ($e^-$), neutrinos $(\bar{\nu}_e)$, $\gamma$-rays ($\gamma$), and neutrons, in addition to reactions that produce alpha particles, and fission fragments. These particles power the associated radioactive heating, and their energy distribution is therefore vital to understanding the details of their thermalization and contribution to the observed light curve \citep{2016ApJ...829..110B,Brethauer:2024zxg,2023MNRAS.520.2558B}. 
The modeling of these emission profiles is itself a challenge, as it not only requires the use of a sophisticated nuclear reaction network, but also appropriate emission spectra for thousands of nuclear species. As a result, kilonova models generally calculate the energy generation rate for each channel via the Q-value of the decay and then make simple assumptions about how that energy is distributed between particles (see e.g. \cite{2021ApJ...910..116K}), although some papers have attempted more detailed models, such as \cite{2023ApJ...954L..41S}, which have incorporated detailed $\gamma$-ray emission spectra, albeit only those experimentally measured.

Recent work \citep{MUMPOWER2025101736} has provided a general purpose framework for calculating the emission spectra for $\beta$-decay, as well as tabulated results for the decay of each nuclear species. In this work, we combine these theoretical spectra calculations with those of a nuclear reaction network to simulate state-of-the-art emission spectra for $r$-process nucleosynthesis. These emission spectra are consistently calculated for all nuclei which undergo $\beta$-decay for all particles: electrons, neutrinos, $\gamma$-rays, and neutrons, and their inclusion in future radiative transport models will enable a more precise interpretation of observations. We organize our paper as follows: In $\S\ref{sec:method}$, we describe the methodology for the calculation of these emission spectra. In $\S\ref{sec:result}$, we present the obtained emission spectra, highlighting the unique features that are observed, and discuss the potential implications for the modeling of kilonova light curves. In $\S\ref{sec:observe}$, we estimate the potential for direct observation of these emission spectra. We summarize and conclude in $\S\ref{sec:conclude}$.

\section{Methodology}
\label{sec:method}
The nuclear reaction for a species undergoing $\beta$-decay can be written as:
\begin{equation}
_Z^AX \rightarrow _{Z+1}^{A}X^{*} + e^- + \bar{\nu}_e
\end{equation}
where A and Z are the mass and atomic number of the decaying nucleus. In addition, nuclei often decay into an excited state of the daughter nuclei, leading to the emission of large numbers of $\gamma$-rays during the de-excitation of this daughter nucleus. Excited states of the daughter can spontaneously emit neutrons, leading to $\beta$-delayed neutron emission:
\begin{equation}
_Z^AX \rightarrow _{Z+1}^{A-C}X^{*} + e^- + \bar{\nu}_e+Cn
\end{equation}
where $C$ is the number of neutrons that are emitted. 
Because the energy release for $\beta$-decays of $r$-process nuclei is normally large ($\sim$ MeV), the energies of the emitted particles can also be substantial. As a consequence, these emitted particles can have significant effects on the modeling of the kilonova, and it is therefore important to understand the distribution of these emitted particles as a function of energy and time.

We define the emission spectrum for species $(i)$ as $S^{(i)}(E,t)$ giving the number of emitted particles per unit time per unit energy per unit mass. This convention is adopted because spectra are often binned in energy, therefore the value of $S^{(i)}(E,t)$ is independent of the bin width. The total number of particles emitted per unit time per unit mass $N^{(i)}(t)$ can be calculated as:
\begin{equation}
\label{eq:numParticles}
N^{(i)}(t)=\sum_j S^{(i)}_j(E,t)w_j,
\end{equation}
where $S_j$ and $w_j$ are the emission spectrum and bin width for the $j^{th}$ energy bin. 
We can calculate the total emission flux by assuming the emitted particles in a given bin are uniformly distributed across the energy range: 
\begin{equation}
\label{eq:Flux}
\Phi^{(i)}(t)=\sum_j S^{(i)}_j(E,t)w_j E_j,
\end{equation}
where $\Phi^{(i)}(t)$ is the emission flux per unit mass, which has units of energy per unit time per unit mass, $w_j$ and $S_j$ are as defined above, and $E_j$ is the average energy of the bin. We can also calculate the average energy of emitted particles as a function of time by dividing the total emission flux by the number of particles:
\begin{equation}
\label{eq:averageE}
\bar{E}^{(i)}(t) = \Phi^{(i)}(t) / N^{(i)}(t) = \frac{\sum_j S^{(i)}_j(E,t)w_j E_j}{\sum_j S^{(i)}_j(E,t)w_j}
\end{equation}

To calculate this emission spectra from $r$-process nucleosynthesis, we shall combine the results from detailed spectra calculations for the $\beta$-decay of individual nuclei with a reaction network which quantifies the number of $\beta$-decays that occur for each nucleus. We therefore factorize the emission spectra as follows:
\begin{equation}
\label{eq:spectra}
S^{(i)}(E,t)= \sum_{^A_ZX} F_{^A_ZX}(t) S^{(i)}_{^A_ZX}(E)
\end{equation}
where $S^{(i)}_{^A_ZX}(E)$ is the spectral emission for species (i) of the nucleus ${^A_ZX}$ per unit decay as a function of energy  (number per unit energy per decay), and $F_{^A_ZX}(t)$ is the reaction flow for the nucleus $^A_ZX$ as a function of time (decays per second per unit mass), and the sum is over every nucleus which decays in the network. We note the importance of the use of reaction flow to characterize the number of decays per unit mass\textemdash while this quantity is in principle just the abundance of the species as a function of time multiplied by its decay rate, in practice, there are many species where the abundance is small but the overall reaction flow and spectral contribution are significant. For a prime example of this, we highlight Tl-208, with a half-life of 3 minutes. It is fed through a long decay chain by Ra-228, with a half-life of 5.75 years, which produces significant spectral contributions even though it is never of significant abundance at such timescales (see \cite{Vassh2024} for more details).

We summarize our methodology for the calculation of each of these two quantities below. 

\subsection{Individual Emission Spectra}
We use the tabulated $\beta$-decay spectra from \cite{MUMPOWER2025101736}, which provides the electron, $\gamma$-ray, neutrino, and neutron spectra for neutron-rich nuclei with A $>$ 8. The emission spectra are calculated using a statistical, multi-phase approach. he calculation first obtains a $\beta$-strength function, giving the distribution of excited daughter nucleus states which the decaying parent initially transforms into. If the nucleus has a level structure and $\beta$-decay scheme which are experimentally known, it is used, otherwise the Quasi-particle Random Phase Approximation (e.g., \cite{KRUMLINDE1984419,1997ADNDT..66..131M,2003PhRvC..67e5802M,PhysRevC.93.025805}) is used to estimate the $\beta$-strength function. From this, the emission spectra of the emitted electron and anti-neutrino are calculated under the assumption that the total available decay energy is divided between them (zero kinetic energy for the daughter nucleus). The shape of the spectra is determined from statistical considerations, where the density of states can be written as:
\begin{equation}
p(Q_\beta)\sim p_e^2 p_\nu^2 F(Z+1,A,E_e) (1-f(E_e))\delta(Q_\beta-E_e-E_v)
\end{equation}
where $p_e$ and $p_\nu$ are the momenta of the electron and antineutrino, respectively, $F$ is the Fermi function for $\beta$-decay, which accounts for distortion in the electron wave function due to the interaction between the nucleus and the electron, and $f$ is the Fermi-Dirac distribution, which accounts for Pauli blocking in the ambient electrons. From the shape of the density of states, the total decay energy can be used to normalize the electron spectrum, and via conservation of energy, the neutrino spectrum. This result is then summed and appropriately weighted over the $\beta$-strength function for possible daughter nucleus states to get the total emission spectra.

The $\gamma$-ray and neutron emissions are calculated from a Hauser-Feshbach statistical approach. For each excited daughter nucleus state which can be accessed from $\beta$-decay, the relative branching ratios of $\gamma$-ray emission (to all possible lower energy states) and neutron emission (to all possible states of the new daughter nucleus) are calculated, creating a web of possible decay paths. These decay paths chain into additional daughter nuclei when $\beta$-delayed neutron emission occurs. From these pathways, the system is treated statistically as a weighted sum over all possible paths, and the emission spectra is the corresponding weighted sum of the spectral emissions of the individual path components. For more details of the mathematical treatment of the spectra calculation, we refer the reader to \cite{MUMPOWER2025101736} and the references therein. 

The frame work used above assumes that beta decays proceed via ``allowed" transitions. First-forbidden transitions are allowed in the calculation of the beta-strength; however, the shape of the spectra is assumed to be the same as allowed transitions. In practice, the assumed shape of the spectra for these transitions would include an additional, energy dependent shape factor which accounts for the orbital angular momentum carried by the leptons. In addition, higher-order transitions are neglected, and if included, would allow different levels to be populated, altering the emission spectra. It is uncertain how much the limitations of our approach will affect the resulting emission spectra. It is suggested that most light-to-mid mass nuclei undergo allowed transitions (see e.g. \cite{PhysRevC.67.055802}); however, it is unclear how much first-forbidden transitions contribute to heavier elements (see e.g. \cite{PhysRevC.93.025805,PhysRevC.102.034326,2025arXiv251102999R}). We emphasize the importance of further development in the calculation of nuclear structure of heavy and superheavy elements in order to accurately quantify the effects on emission spectra.

\subsection{Nuclear Reaction Flow}
We simulate nucleosynthesis with version 1.6.0 of the Portable Routines for Integrated nucleoSynthesis Modeling (PRISM) reaction network \citep{2021PhRvC.104a5803S}. PRISM uses a robust adaptive timestepping algorithm which automatically reduces the timestep if the abundance of any species or the conditions of the environment change too quickly, which fully resolves all species in the network numerically, even those with low abundance, which allows for accurate calculation of the reaction flow.
The nuclear input to PRISM is based on the 2012 version of the Finite Range Droplet Model
\citep{2012PhRvL.108e2501M,2016ADNDT.109....1M}. Radiative capture and fission rates are calculated with the CoH$_3$ statistical Hauser–Feshbach code \citep{2019arXiv190105641K,2021EPJA...57...16K}. 
$\beta$-decay rates, including delayed neutron emission, are calculated assuming statistical de-excitation from excited states \citep{2016PhRvC..94f4317M,2018ApJ...869...14M}. We highlight that these $\beta$-decay rates are calculated from the same assumptions as in the calculation of the spectra above, such that the probability for a nucleus to undergo delayed neutron emission in the reaction network is the same as in the calculation of the spectra.
The remaining reaction rates (e.g. alpha decay and other less substantive reaction types for the $r$-process) are obtained from the REACLIB database \citep{2010ApJS..189..240C}. 
Nuclear fission is handled as in \cite{Vassh2019}. 
\begin{figure}[ht!]
\plotone{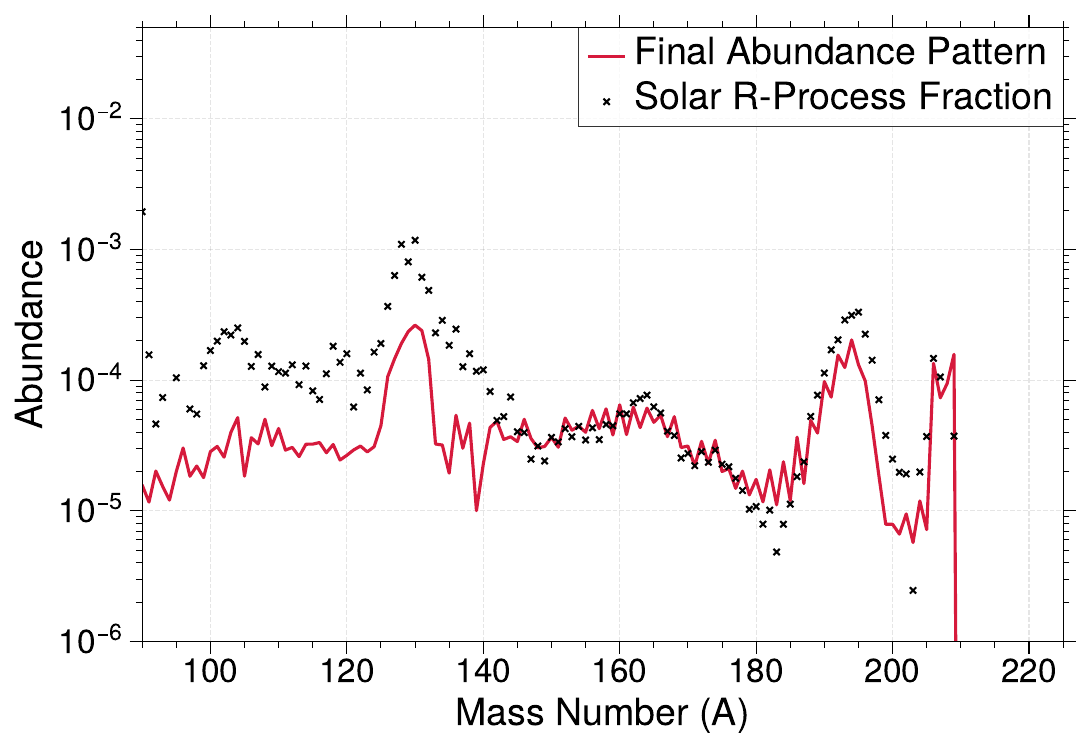}
\caption{Final abundance pattern (at 100 Myr ) of our nucleosynthesis calculation (red) as compared to the solar \textit{r}-process residuals (black). The two patterns are normalized to be identical at A = 150.
\label{fig:5}}
\end{figure}
\begin{figure*}[t]
\epsscale{1.1}
\plotone{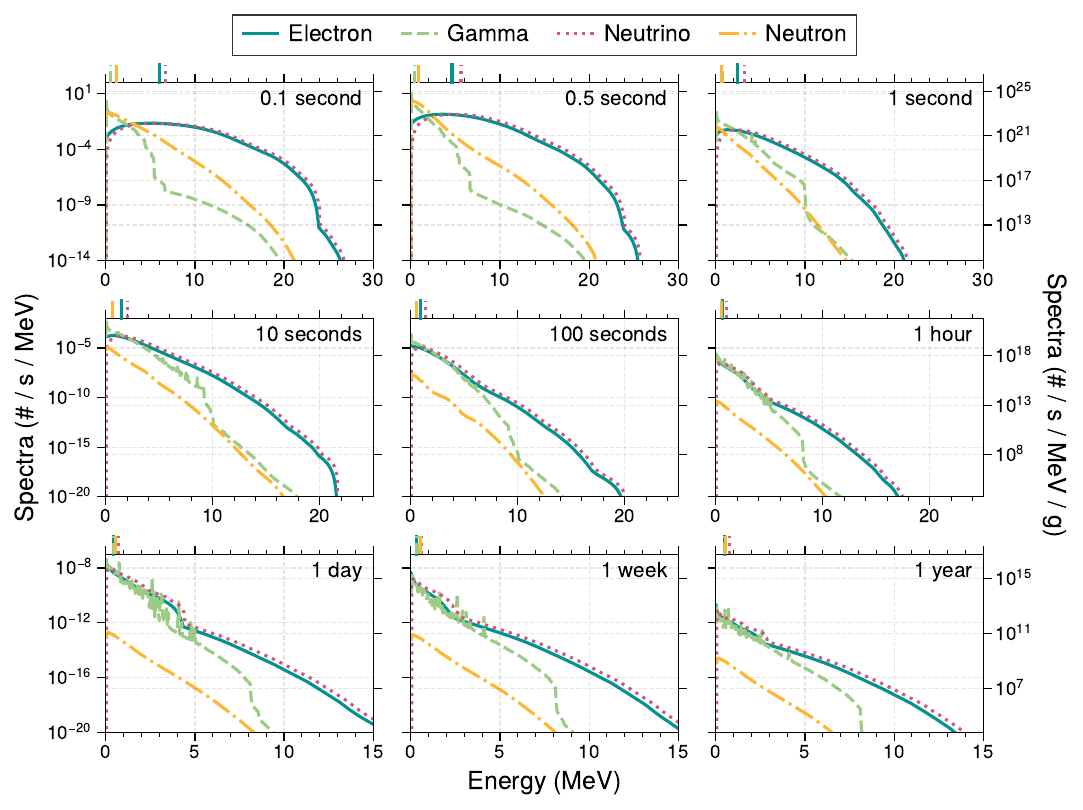}
\caption{Each panel gives the calculated $\beta$-decay spectra for electrons (solid blue), $\gamma$-rays (dashed green), neutrinos (dotted purple), and neutrons (dashdot orange) at the corresponding time, in units of (\#/s/MeV) (per unit abundance) (left axis). The right axis characterizes these spectra scaled to the emission per gram of emitted material. The colored ticks above the plot region give the average energy per particle of the corresponding species at the given time. Note the axes bounds change between rows.
\label{fig:1}}\textit{}
\end{figure*}

To simulate the composition of \textit{r}-process material we use the trajectory (b) of \cite{2025ApJ...982...81M}, which produces a robust $r$-process with substantial actinide production. Fig. \ref{fig:5} shows the final abundances of this trajectory, which are a good match to the solar $r$-process residuals above the second peak of the $r$-process (A $\sim$ 130). 

This trajectory is obtained through modeling the cocoon of a $\gamma$-ray burst, which has been suggested to be a site of the $r$-process due to photo-hadronic interactions in the jet head \citep{2025ApJ...982...81M}. Our choice of this trajectory as opposed to many others from more conventional $r$-process sites which can also produce a robust $r$-process (e.g., \cite{2024ApJ...962...79S}) is due to the accuracy with which the solar residuals are reproduced by this trajectory; however, we do not expect that the  results which we shall discuss below will change significantly with a more conventional choice of trajectory. 
The cocoon is modeled with the density profile of:
\begin{equation}
    \rho(t) = \rho_0 \left( 1 + \frac{t}{\tau_1} + \left(\frac{t}{\tau_2}\right)^\xi \right)^{-1} \ ,
\end{equation}
where $\xi$ = 2, $\tau_1$, $\tau_2$ are characteristic timescales, for which we use $\tau_1 = \tau_2 = 3.5 \times 10^{-2}$ s, and $\rho_0$ is the initial density, for which we use $3.2 \times 10^4$ g/cm$^3$. The temperature is assumed to evolve as an adiabatic gas:
\begin{equation}
    \label{eqn:tempfromrho}
    T(t) = T_0 \left( \frac{\rho(t)}{\rho_0} \right)^{\gamma-1} \ , 
\end{equation}
where $\gamma=4/3$ (radiation dominated), and we take $T_0 = 2$ GK. We take the initial electron fraction to be $Y_e$= 0.034. For more details of the physical motivation and assumptions behind this trajectory, we refer the reader to \citep{2025ApJ...982...81M}. 

\section{Results}
\label{sec:result}
\begin{figure*}[t]
\plotone{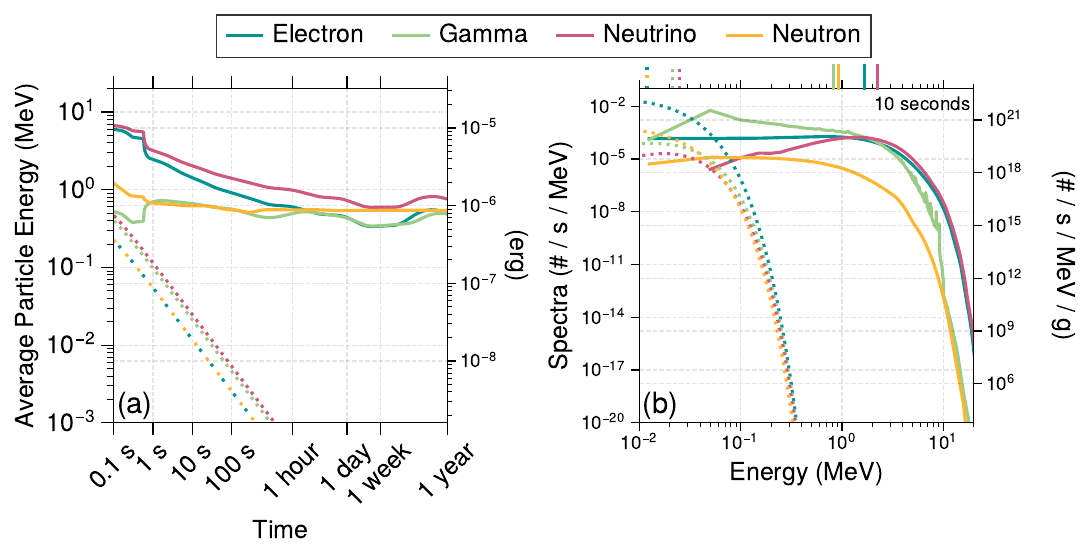}
\caption{\textbf{(a)} Average particle energy as a function of time. The solid curves give the result from the calculation, while the dotted curves give the expected value assuming thermal equilibrium. \textbf{(b)} Spectral energy distribution at 10 seconds (solid curves) as compared to the expected distribution assuming thermal equilibrium (dotted curves). The colored ticks on the upper x-axis give the average energy of the emitted particles (solid) and that expected from the a thermal distribution (dotted).
\label{fig:3}}
\end{figure*}
In Fig. \ref{fig:1}, we show the particle emission spectra for electrons, $\gamma$-rays, neutrinos, and neutrons as a function of energy at several different times. 
We note the following conclusions.

The electron and neutrino spectra are almost identical. In the first second of emission, their emission profiles are significant to very high energies ($\sim$ 10-20 MeV). For the first 10-  /100 seconds, this trend continues, with significant production up to $\sim$ 5 MeV with an average energy of $\sim 1$ MeV, and then a gradual but persistent falloff to $\sim$ 20 MeV. We note that interesting features in the distribution are consistently present, but not constant as a function of time, such as the bump around $\sim$10 MeV at 100 seconds.   On longer time scales, the spectra remains broadly distributed in energy, though the distribution generally continues to decrease in the number of higher-than-average energy particles. We attribute the smooth high energy tail in the spectrum which is seen up to long time scales to the decays of the byproducts of fission, which produce a consistent shower of a wide distribution of particles to long time scales.  In general, the neutrino spectrum is slightly higher average energy ($\sim 15\%$) than that of the electron. The consistency in this trend that is observed as well as the similarity of the shape of the neutrino and electron spectra is likely a consequence of the statistical approach taken to partition the available energy between the neutrino and electron. 

The $\gamma$-ray spectrum is characteristically different from the electron and neutrino spectra. Initially, the distribution is bimodal in nature, with a strong peak at low energy ($\sim$ 1 MeV), then a sharp fall-off, followed by a flattening to $\sim$ 15-20 MeV. This rapidly morphs into a unimodal distribution by $\sim$ 1 second. Initially, the distribution is generally more smooth, with not many specific spectral lines to be identified. This is attributed to the lack of available nuclear data for the many short-lived nuclei far from stability which decay at early times. This lack of data means that there are not specific spectral features, and the $\gamma$-ray spectra is determined from theoretical calculations. As time increases ($\gtrapprox$ 1 hour), we begin to see the emergence of the specific spectral lines of longer-lived nuclei. These come to dominate the spectra by $\sim$ 1 day. These spectral lines can be easily paired with specific nuclei, for instance with the decay of $^{208}$Tl as proposed by \cite{Vassh2024}, which is responsible for the 2.6 MeV line, and $^{140}$La, which is responsible for the 2.0, 3.1, and 3.3 MeV lines seen at 1 week, although radiation transport may be necessary to determine the observability of the lines at earlier times. We expect that additional $\gamma$-rays are emitted during prompt fission emission, and will provide a larger high energy tail to the distribution (see e.g. \cite{2020ApJ...903L...3W}), but this is not included in the calculation, which only considers the $\beta$-decay spectra.

The neutron spectrum is the most featureless of the spectra. Initially, this spectrum is much more significant than that of the $\gamma$-ray spectra, but shrinks to become comparable to the $\gamma$-ray spectra by $\sim$10-100 s and then significantly smaller by later times. We attribute this to the fact that while neutron capture is still occurring (~$\lessapprox$ 100 s), the nuclei which are created are farther from equilibrium and therefore have a much larger number of $\beta$-delayed neutron emission. Indeed, as time increases, the $\gamma$-ray spectrum grows to match and surpass the neutron spectrum (at around $\sim$ 1 second); however, the two remain in close proximity until neutron capture ends ($\sim$ 100 seconds). Beyond 100 seconds, neutrons emitted can be attributed to decays of the byproducts of fission, and like the $\gamma$-rays, it is expected that additional high-energy neutrons are emitted during prompt fission emission, but this is not included in the calculation.

The solid lines of Fig. \ref{fig:3}a indicate the average particle energy (Eq. \ref{eq:averageE}) for each particle as a function of time. 
In general, we note the average energies of the electron and neutrino are high for the first several seconds: $\sim 5$ MeV until $~1$ second, then 1.5-2.5 MeV until $\sim 10$ seconds, not dropping below 1 MeV until neutron capture ends ($\sim 100$ seconds). In contrast, the average energies of neutrons and $\gamma$-rays remain lower the entire time ($\lessapprox$ 1 MeV). However, for all species, the average energy is substantially higher than what would be expected from thermal equilibrium (dotted lines).
To further illustrate the highly non-thermal nature of the particle spectra, we compare in Fig. \ref{fig:3}b the spectral distribution at $t=10$ seconds to what would be expected from a thermal distribution. The difference is striking, showing a much harder spectrum than if thermal equilibrium was assumed. 

\begin{figure}
\epsscale{1.1}
\plotone{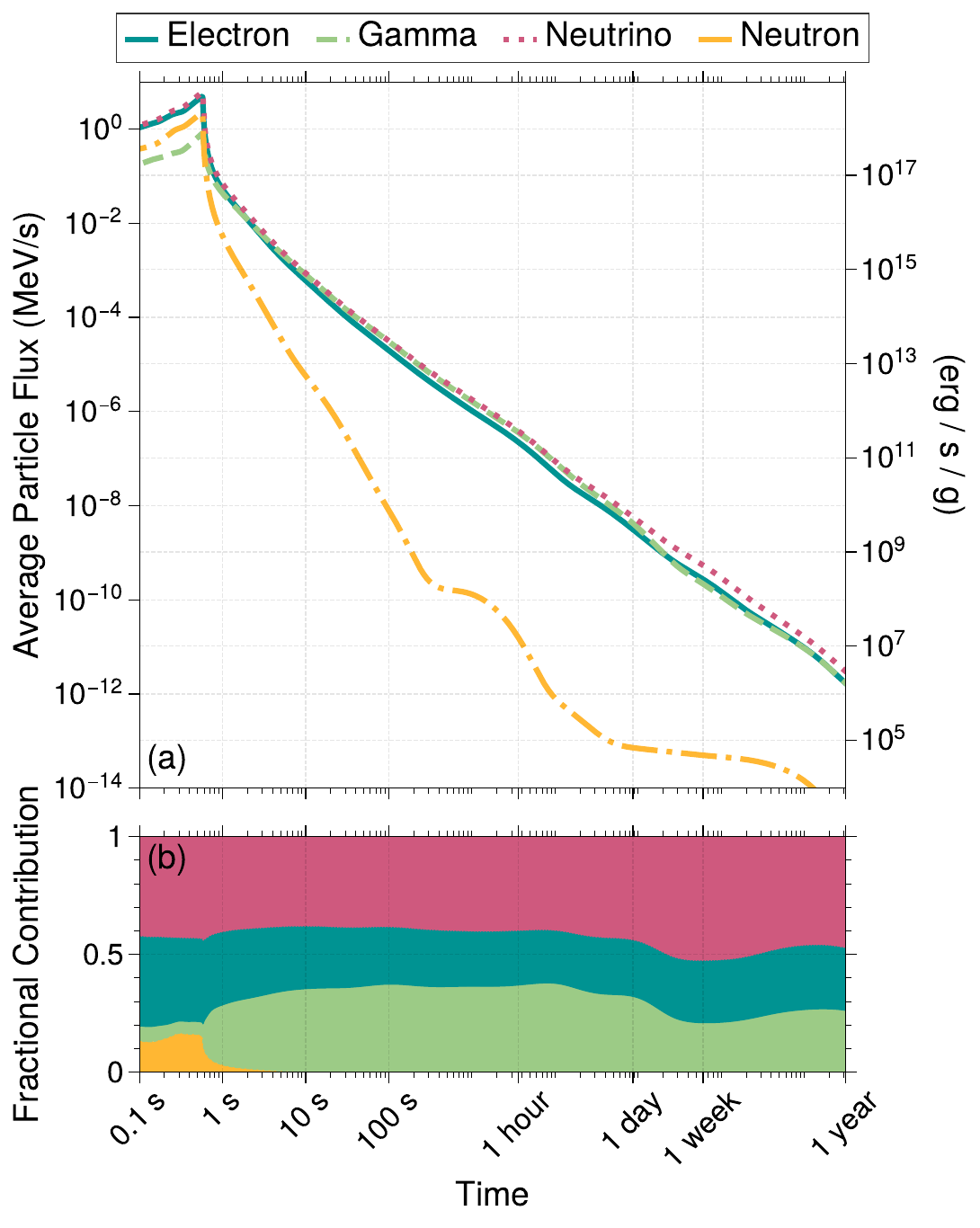}
\caption{\textbf{(a)} Total energy (flux) of emitted particles in units of (MeV/s) per unit abundance (left axis) as a function of time. The right axis gives the emission scaled to the conventional units of (erg/s/g). \textbf{(b)} The relative contributions of each channel to the total emission energy.
\label{fig:2}}
\end{figure}

Fig. \ref{fig:2}a gives the total energy flux (Eq. \ref{eq:Flux}) as a function of time, with the relative contributions from each channel shown in \ref{fig:2}b. We note that the distribution of available energy is neither equal between channels nor constant in time. The largest fraction of energy goes to the neutrinos (between $\sim 40\%$ and $50\%$), while the rest is mainly divided between the $\gamma$-ray and electron emission. At early times, ($\lessapprox 1$ s), the contribution to $\gamma$-rays is small, and this is the only time period in which the contribution to neutrons is significant, though most of the difference in $\gamma$-rays is allocated to an increased electron contribution. 

As the energy distributions of the emission spectra are an important input for the associated kilonova modeling, both the non-thermal distributions and the unequal allocation of available energy between channels may have significant impact. 
Full radiative transport is necessary to estimate the timescale for the emission spectra to become thermalized and to understand the effects on the resultant light curve. We emphasize that our calculations have no radiative transport, but are intended to provide self-consistent modeling of particle spectra which can be incorporated into the corresponding transport codes to enable improved understanding of kilonova light curves.

The particle emission spectra underlying the results presented are publicly available via Zenodo \citep{gross2025dataset}.

\section{Observability of Signals}
\label{sec:observe}
\begin{figure*}[t]
\plotone{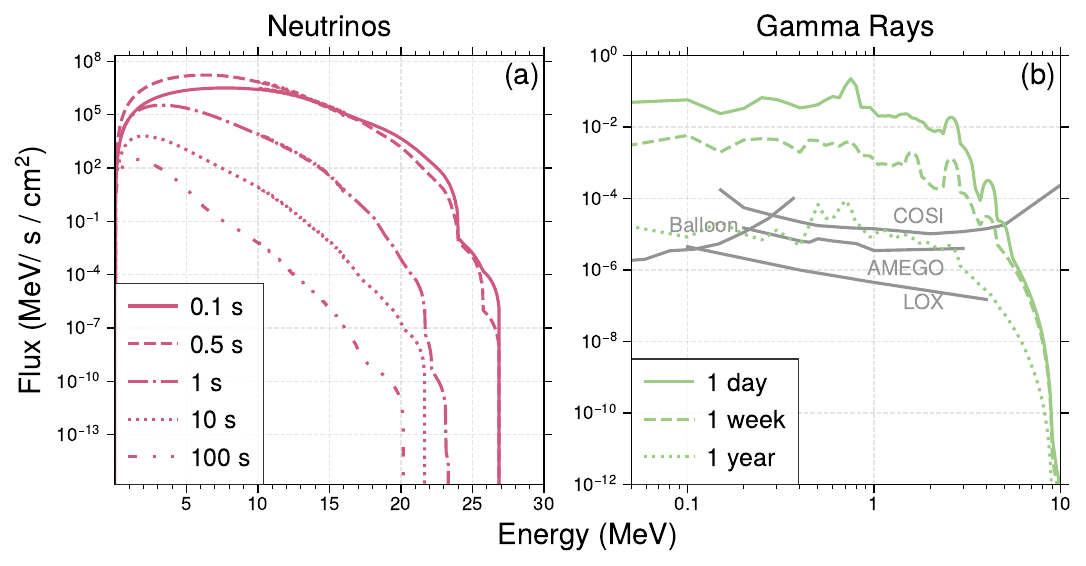}
\caption{\textbf{(a)} Neutrino flux at different times as a function energy assuming an ejecta mass of $0.01M_\odot$ at a source distance of 15 kpc. \textbf{(b)} $\gamma$-ray flux at different times as a function of energy assuming an ejecta mass of $0.01M_\odot$ at a source distance of 15 kpc. This spectrum was broadened as described in Eq. \ref{eq:broaden} with an assumed expansion velocity of $0.1c$. The gray lines show the detection limits of next-generation $\gamma$-ray detectors for a 1 Ms ($10^6$ s) duration exposure. \label{fig:6}}
\end{figure*}

While these spectral emission profiles are an important component of the modeling of kilonova with significant impact on the final light curve, we can also consider the possibility of the direct observation of these emission profiles. Emitted electrons and neutrons will interact with the surrounding medium and will not be directly observable. However, emitted neutrinos will escape, and if they are produced in sufficient numbers at early times, they may be observable. $\gamma$-rays are initially opaque to observations, but at later times, the opacity will lower and some $\gamma$-rays will escape. The timescale for which $\gamma$-rays in the ejecta become less opaque is estimated be of order hours to days (see e.g. \cite{2024MNRAS.533..994G}, which estimate the a timescale of $\sim$ 1 day for a GW180817-like ejecta). Full transport calculations are required to model the ejecta to determine the expected spectral signatures at early times, but at later times, if the emitted $\gamma$-rays are produced in sufficient number, specific emission lines may be directly observable. In this section, we provide an estimate of the neutrino and $\gamma$-ray flux resulting from $r$-process emission.

We assume that the emitted particles are a uniform, point-like emission. We can calculate the total incident flux from a total ejecta mass M at a distance D from the source:
\begin{equation}
\label{eq:IncidentFlux}
F^{(i)}(E,t^*)= \Phi^{(i)}(E,t) (\frac{M_{ejecta}}{M_n})\frac{1}{4\pi D^2},
\end{equation}
where $\Phi^{(i)}(E,t)$ is the total flux of species (i) rays emitted per second per unit energy (per unit abundance), similar to that in Eq. \ref{eq:Flux}, $M_n$ is the average nucleon mass ($1.67 \times 10^{-24}$ g), $M_{ejecta}$ is the assumed total mass which undergoes nucleosynthesis, and $D$ is the distance to the source.
We nominally take $D$ = 15 kpc and $M_{ejecta} = 0.01 M_{\odot}$ as conservative but reasonable values for an event which occurs in our galaxy. 
\subsection{Neutrinos}
We display our results for the incident neutrino flux in Fig.~\ref{fig:6}a at varying times over the first 100 seconds. 
To estimate the possibility of detection, we consider only the first 10 seconds of emission, which constitutes 95\% of the total neutrino emission. Following the approach of \cite{2023PhRvD.108l3038A}, we consider inverse $\beta$-decay as the detection channel. The number of events in a Cherenkov detector can be estimated as (e.g. \cite{2023PhRvD.108l3038A}):
\begin{equation}
\mathcal{N} \approx N_p \bar{\sigma}_{\text{IBD}} F_{total}^{\nu}(t)
\end{equation}
where $\bar{\sigma}_{\text{IBD}} \approx 9.5 \times 10^{-42} \ cm^2\times [\bar{E}^{\nu} / (10 \text{ MeV})]^2$, and $N_p \sim 1.3 \times 10^{34} \times [M_T / (200 \text{ kton})]$, $\bar{E}^\nu$ is the average neutrino energy, $M_T$ is the detector size, and $F^\nu_{total}(t)$ is the total number of incident neutrinos by time t, which we estimate from the integration of Eq. \ref{eq:IncidentFlux} to be $F^\nu_{total}(10 \  s) = 1.54 \times 10^8$ cm$^{-2}$. We adopt $\bar{E}^\nu$ = 3 MeV and $M_T$ = 260 kton, which corresponds to the approximate size of Hyper-Kamiokande. With these values, we find $\mathcal{N} \sim 2.2 $ observed events. Therefore, $r$-process nucleosynthesis events in our galaxy may be observable, though we emphasize that this only an estimate which ignores many important components of a thorough analysis, such as oscillations, detector response, etc.

\subsection{$\gamma$-Rays}
The $\gamma$-rays will undergo spectral broadening due to the expansion of the ejecta. Assuming a spherical mass distribution for the ejecta with homologous expansion at velocity $v$, the resultant distribution of a spectral line of energy $E_0$ can be written as (See \S\ref{sec:broaden} for a derivation):
\begin{equation}
\frac{dN}{dE}=\frac{3N_{total}E_0}{4E^2v}(1-\frac{(1-E_0/E)^2}{v^2})
\label{eq:broaden}
\end{equation}
where $N_{total}$ is the number of particles emitted per unit time, and the range of possible energies E (for a line of energy $E_0$) are $E_0/(1+v)\leq E\leq E_0/(1-v)$, where $v$ is in units of $c$. Taking a nominal value of $v = 0.1\ c$, and concatenating the above spectral broadening with Eq.~\ref{eq:IncidentFlux}, we calculate and display the incident $\gamma$-ray flux for an ejecta of mass $0.01 M_\odot$ at distance 15 kpc in Fig.~\ref{fig:6}b. We compare this to the predicted observational capabilities of existing and future $\gamma$-ray detectors. Our results suggest that $\gamma$-rays from events within our galaxy should be observable for weeks to months. We note that this is a conservative estimate: the ejecta mass was taken to be only $0.01 M_\odot$, while some models suggest that the ejection mass could be as high as $0.1M_\odot$. Our results are comparable with those of other studies which perform detailed radiation transport (see e.g. \cite{2025arXiv250317445J}, \cite{10.1093/mnras/staf1147}), though we highlight the difference in focus \textemdash while these studies focus on the observability at early times out to large distances (40 Mpc), we wish to highlight the direct observability of the emission profiles for more modest distances (15 kpc) at later times after the opacity has lowered.
\section{Conclusions}
\label{sec:conclude}

We have presented a first-principles calculation of the emission spectra of electrons, $\gamma$-rays, neutrinos, and neutrons as a function of energy and time for $r$-process nucleosynthesis. The spectra that we calculate have several unique properties: they are decisively non-thermal, with average energies far higher ($\gtrapprox$ 1 MeV) than would be expected from a thermal distribution ($\sim$ keV). We also find that the partitioning of energy is not even between channels as a function of time. 
For the majority of the evolution, the largest fraction of the energy is partitioned to neutrinos, with smaller but significant amounts going to electrons and $\gamma$-rays, and very little to neutrons, except at early times. Our microscopic calculations provide the spectral detail necessary for higher fidelity kilonova modeling, and their inclusion in future radiative transport models will enable a more precise interpretation of observations. 

We also estimate the observability of the produced $\gamma$-ray and neutrino signal, finding that $r$-process activity resultant from GRBs may be directly observable within our galaxy for significant nucleosynthesis $(M \gtrapprox 0.01 M_\odot$). The possibility of direct observation offers an exciting opportunity to bypass the degeneracies of kilonova modeling and probe nucleosynthesis through other messengers. In particular, spectral features of the $\gamma$-rays can be tied directly to the specific nuclei (such as the 2.6 MeV line produced by the $\beta$-decay of $^{208}$Tl). More work is necessary to identify and characterize these observational features in a variety of nucleosynthetic scenarios.
\\
\\
\\

\begin{acknowledgments}

\noindent LANL is operated by Triad National Security, LLC, for the National Nuclear Security Administration of U.S. Department of Energy (Contract No. 89233218CNA000001). 
Research presented in this article was supported by the Laboratory Directed Research and Development program of Los Alamos National Laboratory under project number 20230052ER.

\end{acknowledgments}

\bibliography{refs}{}
\bibliographystyle{aasjournalv7}
\appendix
\section{Doppler Broadening of Line Profiles for Ejecta}
\label{sec:broaden}
We model our ejecta as a spherically symmetric mass distribution which is homologously expanding. The ejecta is emitting photons at a rate $\dot{\epsilon}$ per unit mass per second with wavelength $\lambda$. Homology implies that the velocity profile is proportional to the radius, and we can use this to write the mass profile in terms of the velocity:
\begin{equation}
\frac{dm}{dv}=m_{ex}\frac{3v^2}{v_{ex}^2},
\end{equation}
where $m_{ex}$ is the total ejection mass and $v_{ex}$ is the velocity of the edge of the distribution. 

The Doppler shift due to expansion can be written as:
\begin{equation}
\label{eq:lambdavz}
\lambda(v_z)=\lambda_0(1-v_z),
\end{equation}
where $v_z$ is in units of $c$ and gives the velocity component along the line of sight. We note that since we have homologous expansion, the value of $v_z$ is solely based on the distance from the center of the source along this line of sight. Therefore, we consider the circular slice which is centered along the line of sight with velocity component $v_z$, and can calculate the differential mass of this slice:
\begin{equation}
dm = \frac{3m_{ex} \dot\epsilon}{4v_{ex}^3}(v_{ex}^2-v_z^2)dv_z
\end{equation}
The total photon emission can then be expressed as:
\begin{equation}
\frac{dN}{d\lambda}=\frac{dm}{d\lambda}\dot\epsilon = \frac{3m_{ex} \dot\epsilon}{4v_{ex}^3}(v_{ex}^2-v_z^2)\frac{dv_z}{d\lambda}
\end{equation}
Noting that we have $dv_z/d\lambda=-1/\lambda_0$, and using Eq.~\ref{eq:lambdavz}, we find:
\begin{equation}
\frac{dN}{d\lambda}=\frac{3m_{ex}\dot\epsilon}{4v_{ex}\lambda_0}\left(1-\frac{(1-\lambda/\lambda_0)^2}{v_{ex}^2}\right)
\end{equation}
where the bounds on this distribution are $\lambda_0(1-v)\leq \lambda \leq \lambda_0(1+v)$. Translating this into the energy regime, we find:
\begin{equation}
\frac{dN}{dE}=\frac{3N_{total}E_0}{4E^2v_{ex}}(1-\frac{(1-E_0/E)^2}{v_{ex}^2}),
\label{eq:broaden}
\end{equation}
where $N_{total}=m_{ex}\dot\epsilon$ is the total emission rate for the entire ejecta, and the bounds on the distribution are now $E_0/(1+v)\leq E \leq E_0/(1-v)$.
\end{document}